\documentclass[twocolumn,prl,aps,floatfix,superscriptaddress,amssymb]{revtex4}

\usepackage{graphicx}
\usepackage{subfig}
\usepackage{epsfig}
\usepackage{amsmath}
\usepackage{textcomp}
\usepackage[justification=centerlast]{caption}


\begin{document}

\title[Orbital Magnetization of Quantum Spin Hall Insulator Nanoparticles]
  {Orbital Magnetization of Quantum Spin Hall Insulator Nanoparticles}

\author{P. Potasz}
\affiliation{International Iberian Nanotechnology Laboratory (INL), Av. Mestre Jos\'e Veiga, 4715-330 Braga, Portugal}
\affiliation{Department of Theoretical Physics, Wroclaw University of Technology, Wybrzeze Wyspianskiego 27, 50-370 Wroclaw, Poland}
\author{J. Fern\'andez-Rossier}
\affiliation{International Iberian Nanotechnology Laboratory (INL),
Av. Mestre Jos\'e Veiga, 4715-330 Braga, Portugal}

\begin{abstract}
Both spin and orbital degrees of freedom contribute to the  magnetic moment of isolated atoms. However,  when inserted in crystals,  atomic  orbital moments are quenched because of the lack of  rotational symmetry  that protects them when  isolated.  Thus, the dominant contribution to the magnetization of magnetic materials comes from electronic spin. Here we  show that 
nanoislands of quantum spin Hall insulators can host robust orbital edge magnetism whenever their
highest occupied Kramers doublet is singly occupied, upgrading the   
spin edge current  into  a charge current.  The resulting 
   orbital magnetization scales linearly with size,  outweighing the spin contribution for islands of a few nm in size. 
   This linear scaling is specific  of  the 
Dirac edge states and very  different from  Schrodinger electrons in quantum rings.
Modelling Bi(111) flakes,  whose edge states have been recently observed,  we show that 
 orbital magnetization is robust with respect to disorder, thermal agitation,  shape of the island and crystallographic direction of the edges, reflecting its topological protection. 
\end{abstract}

\maketitle

A central notion in magnetism is the fact that orbital moments associated to circulating currents  are fragile.  They naturally occur in open-shell isolated atoms\cite{Ashcroft+76}, but these atomic  orbital moments  quench as soon as the atom is placed in a crystal.  Circulating currents in artificially patterned mesoscopic quantum rings\cite{Saminadayar+04}, studied in the last three decades\cite{Buttiker+83,vonOpen+09}, require very special conditions to survive, such as  very low temperatures so that the electrons keep their phase coherence around the entire ring, and small disorder, so that  electrons do not localize.  
In contrast, robust spin currents occur naturally at the edge of quantum Spin Hall insulators (QSHI)\cite{Kane+05,Kane2+05,Bernevig+06} and are topologically protected. 
These spin currents are associated to Kramers doublets, where each state has  a net charge current flowing  with opposite chirality. In a finite sample, these counter-propagating currents can be  associated to magnetic moments with opposite sign for each state in the Kramers doublet.   Since these states are equally occupied, the resulting net orbital moment vanishes. 
Having an insulating bulk and robust spin currents at the edges, QSHI are natural quantum rings\cite{bookrings} for spin currents. 
 The central idea of this paper is  that,  in the case of QSHI nanoislads\cite{Hawrylak+14} (or flakes) with a discrete edge state spectrum,  
 it is possible to turn these robust spin currents into robust charge currents that result in very large orbital moments.
   To do so, two conditions are  sufficient:  a magnetic field has to split the Kramers doublets and,  using electrical gating or chemical doping, only one electron has to occupy  the highest occupied Kramers pair, providing thereby  
  a net edge current, and a large  orbital magnetization. 
 
 Several  systems have been predicted to be QSHI\cite{Hasan+10,Zhang+10} and strong experimental evidence exists that
 CdTe/HgTe quantum wells\cite{Konig+07}, InSb/GaAs quantum wells\cite{Knez+Du+11}, and  with  Bismuth (111) monolayers\cite{Liu+11,Yang+12, Sabater+13,Drozdov+14} host spin filtered edge states essential for our proposal. To substantiate our claim, we choose Bismuth for two reasons:   the topological edge states of  nanoislands of Bi(111) have been recently observed by means of Scanning Tunneling spectroscopy (STM)\cite{Drozdov+14,Kawakami+15},  and a very well tested tight-binding Hamiltonian\cite{Liu+95} is available that makes it possible to compute the electronic structure of systems with thousands of atoms.
   
\begin{figure*}
\epsfig{file=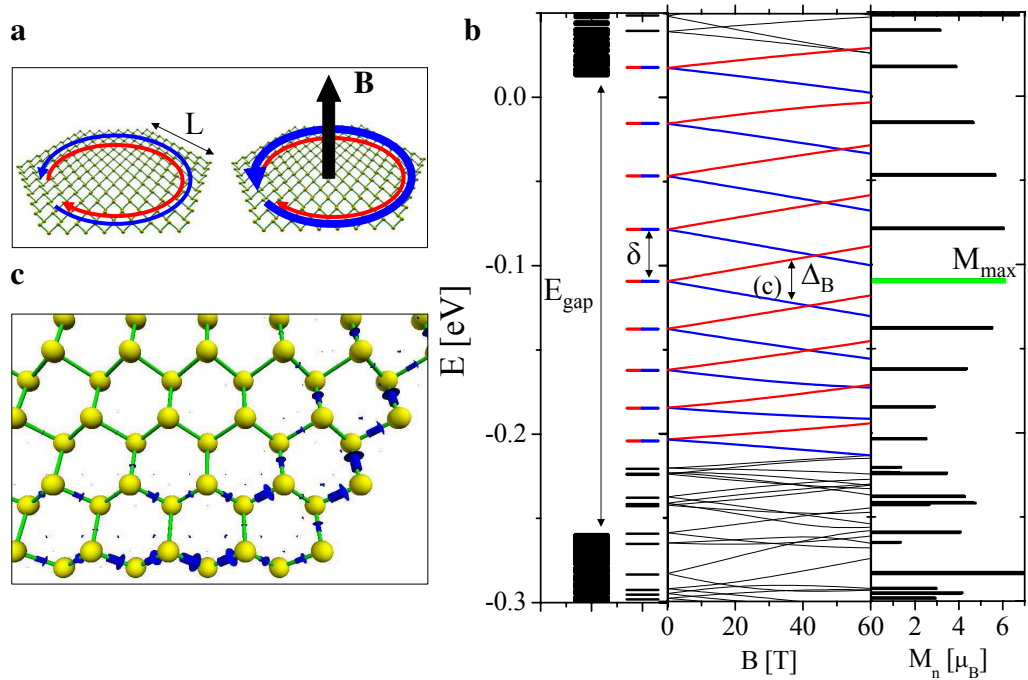,width=5.5in}
\caption{\label{fig:Fig1}{\bf Orbital magnetization of Bi(111) nanoisland edge states}
  {\bf a}  Scheme of edge currents of a given Kramers doublet.  At zero field,  the occupation of both states with opposite orbital magnetization is the same (red and blue arrows). Application of a magnetic field, plus single occupancy of a Kramers doublet, results in net orbital magnetization (thick blue arrow).  {\bf b} Calculated energy spectra of a the 2D Bi(111) bilayer and a flake  with edge length $L\simeq 3.6$nm (N=384 atoms) and evolution of flake spectrum as function of a magnetic field. The bilayer gap $E_{gap}$, energy level spacing  $\delta$, and the  splitting of the Kramers doublets  in a magnetic field $\Delta_{B}$ are indicated. 
  The corresponding magnetic moments, $M_n=\left|\frac{\partial E_n}{\partial B}\right|$   are shown on the right. The largest  magnetic moment $M_{max}$ 
  is highlighted in  green. {\bf c} Calculated local current density (blue arrows) flowing along edges for the state generating $M_{max}$.}
\end{figure*}

Bi(111) bilayer (BL) is a buckled 2D honeycomb crystal (see figure \ref{fig:Fig1}). We model the  Bi(111) nanostructures  with  the same tight-binding model\cite{Liu+95} employed by  Murakami\cite{Murakami+06} to predict that Bi(111) would be QSHI. The same approach has also been used by Drozdov {\it et al.}\cite{Drozdov+14} and by Sabater {\it et al.}\cite{Sabater+13} to model their experimental results.
The Liu-Allen tight-binding model describes Bi with four orbitals ($s,p_x,p_y,p_z$) per atom, with interatomic hoppings up to third neighbors, parametrized with the Slater-Koster approach\cite{Slater+54} and atomic spin-orbit coupling $\lambda \vec{L}\cdot\vec{S}$.  Within this model Bi(111) bilayers naturally come as QSHI\cite{Murakami+06}, with spin-filtered edge states and a gapped  bulk.  The effect of the magnetic field is incorporated by using Peierls substitution\cite{Peierls+33}, with an extra phase accumulated by electron going from site $i$ to $j$, $\varphi_{ij}=2\pi\frac{e}{hc}\int_{r_i}^{r_j}{\bf{Adl}}$, where  ${\bf{A}}$ is the vector potential, $\phi_0=\frac{hc}{e}$ is magnetic flux quantum.

 The electronic structure of a Bi(111)  nanoisland, with hexagonal shape   with six zigzag edges with length  $L_{\rm edge}=3.6$nm each,  is shown in Fig.  \ref{fig:Fig1}a. In panel \ref{fig:Fig1}b we show the energy levels corresponding to the 2D material, to mark the gap of 0.25eV, side by side with the discrete energy spectrum of the island.   We denote the energy difference between adjacent Kramers doublets by $\delta$. Upon application of a magnetic field perpendicular to the island, the in-gap  Kramers doublets split following straight lines, indicated by red (blue) colors for states with increasing (decreasing) energies. Red and blue color lines distinguish the in-gap edge states rotating clockwise and counterclockwise, respectively. The splitting, denoted by  $\Delta_B$, exceeds by far the spin Zeeman splitting,  which actually is not  included in the Hamiltonian. This  indicates that these states carry an orbital moment. Its origin becomes apparent upon inspection of the plot of the current density associated to the one of two states of a Kramers doublet, displayed in figure \ref{fig:Fig1}c, that shows the circulating edge current. It must be noted that some states not in the gap also have large orbital moments. We have verified that their wave functions are not fully localized at the edge, so that the emergence of the large orbital moment could have a different origin.  
 
\begin{figure*}
\epsfig{file=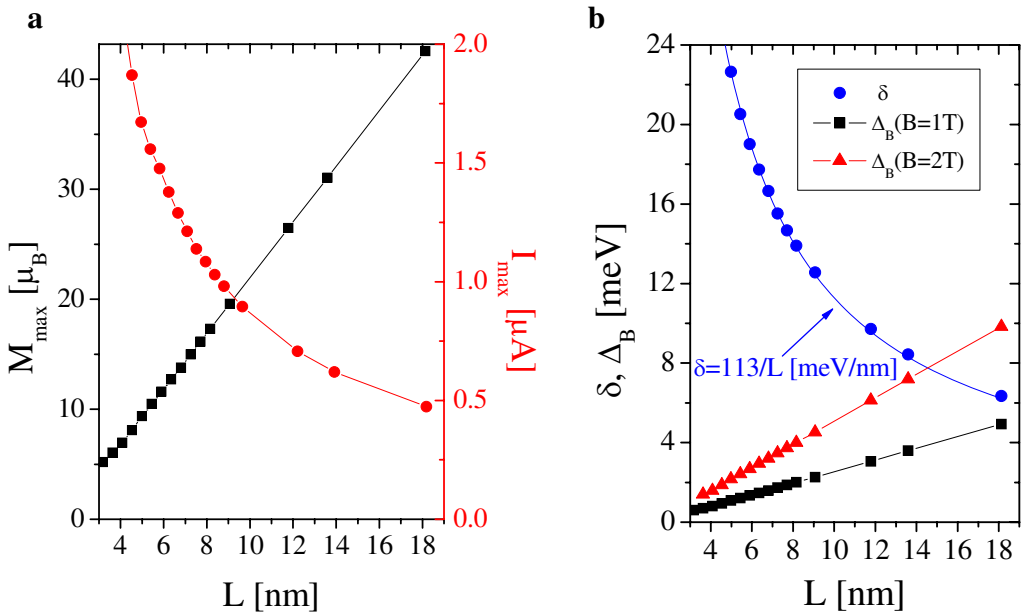,width=5.5in}
\caption{\label{fig:Fig2}{\bf Size scaling analysis.} {\bf a} Maximal orbital magnetic moments $M_{max}$ (see Fig. \ref{fig:Fig2}b) in units of Bohr magnetons $\mu_B$ (black squares) and corresponding current amplitudes $I_{max}$ (red circles) as a function of edge length $L$ of the flake. A linear dependence of orbital magnetic moments as a function of edge length $L$ is clearly seen, with $\frac{\Delta M_{max}}{\Delta L}\approx 2.5\frac{\mu_B}{\rm nm}$. {\bf b} A comparison between characteristic energy scales in the systems as a function of edge length $L$. Energy level splitting of the edge states in the absence of the magnetic field $\delta$ (blue circles) and a Kramers degeneracy splitting in a magnetic field $\Delta_B$ for a magnetic field $B=1$T (black squares) and $B=2$T (red triangles). $\delta=113 meV/L(nm)$ (blue curve) obtained from a fitting procedure, which is characteristic size-dependent quantization ($\propto L^{-1}$) for Dirac Fermions.}  
\end{figure*}

 To be more quantitative, we use the definition of  magnetic moment associated to a given quantum state\cite{Ashcroft+76}
\begin{eqnarray}
M_n=-\frac{\partial E_n}{\partial B}.       
\label{OrbMom}
\end{eqnarray}
The absolute value of the corresponding orbital magnetic moment associated to the in-gap Kramers doublets is shown in \ref{fig:Fig1}b.  In Fig. \ref{fig:Fig2} we show the magnetic moment for the  in-gap  state with the largest $M_n$ for a given island, denoted as $M_{\rm max}$,  
as a function of the island size $L$. The magnitude of $M_{\rm max}$ scales linearly with $L$,  reaching values as high as 
 42 Bohr magnetons ($\mu_B$) for a hexagonal island with $L=18$nm, much higher than the spin contribution $(1\mu_B)$.    For   Schr\"odinger particles in a ring,   the magnetic moment is given by  $M_{\rm Schro}= \mu_B L_z$, where $L_z$ is the azimuthal quantum number,  independent of the size.  In contrast, from the exact solution of a massless Dirac particle moving in  a ring\cite{Peeters+10,Ghosh+13}, equation (\ref{OrbMom}) gives that the magnetic moment of Dirac particles also scales  linearly with the circle radius.  
   Therefore, our results can be interpreted as if the edge states were described by Dirac particles confined in a ring. We can also relate the magnetic moment $M_{\rm max}$ to a persistent edge current through the classical definition of magnetic moment in a loop $M_{\rm max}= I A$, where $A$ is the area of the nanoisland. The current so calculated are shown in Fig. \ref{fig:Fig2}a, and can reach values of 0.5$\mu$A for $L=18$nm.  
 
\begin{figure*}
\epsfig{file=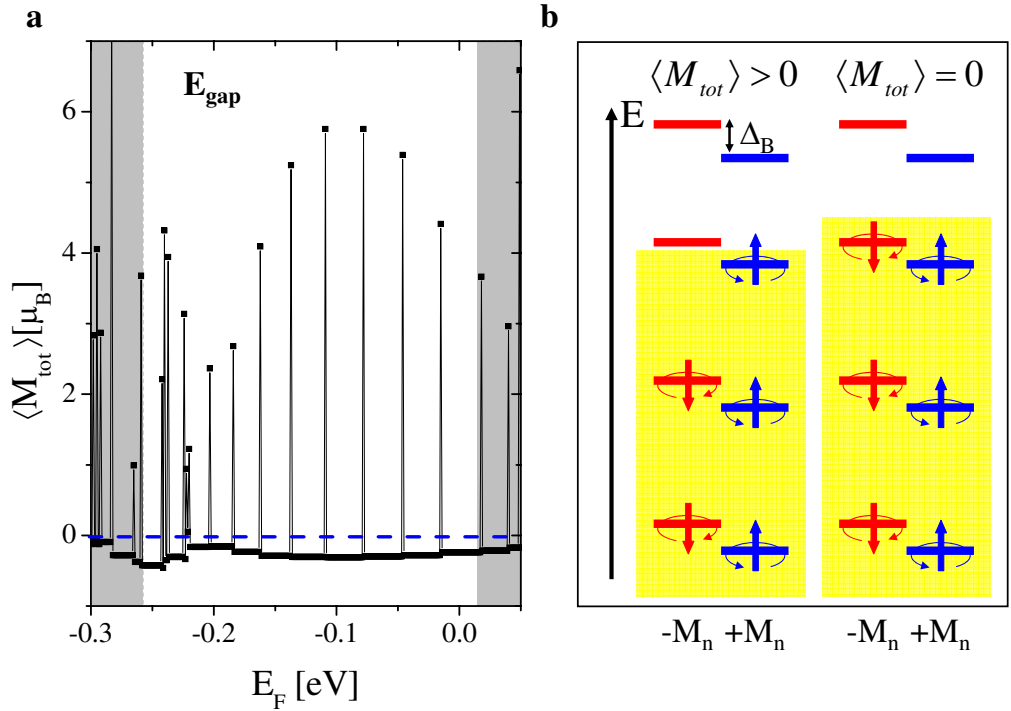,width=5.5in}
\caption{\label{fig:Fig3}{\bf Total magnetization.} {\bf a} Total magnetization $M_{tot}$ as a function of Fermi energy $E_F$ at temperature $T=0$ for the island  with edge length $L\simeq 3.6$nm considered in Fig. \ref{fig:Fig1}. The off-gap states, as defined by the bilayer spectrum, are shown with a grey background.  As the Fermi energy moves in-gap,  total magnetization oscillates . {\bf b} A schematic picture explaining magnetization oscillations shown in a. Red and blue bars correspond to energy levels from a given Kramers doublet split by a magnetic field. Filled  states (yellow area) are occupied by electrons (indicated by arrows) in edge states. When  number of electrons is odd, only one of the states from the highest Kramers doublet is occupied (left panel), inducing orbital magnetic moment $+M$ that contributes to total magnetization. In this state, addition or removal of a single electron results in a quenching of the edge magnetization.}
\end{figure*}

Whereas quantum transport experiments could probe the magnetic moment of individual states\cite{McEuen+04,McEuen+08}, 
magnetometry  experiments are sensitive to the  total magnetization, which involves contributions
from all occupied states:
\begin{eqnarray} 
\langle M_{\rm tot}\rangle=\sum_{n} f(\epsilon_n) M_n, 
\label{Magnet}
\end{eqnarray}
where the sum runs over all the state of the island and $f(\epsilon_n)$ is occupation of the individual states. 
In Fig. \ref{fig:Fig3}a we show the  $T=0$ magnetization  as a function of Fermi energy $E_F$ of the island considered in Fig. \ref{fig:Fig1}, for a magnetic field of $B=1$T. It is apparent that, whenever a split Kramers doublet is singly occupied (as seen in figure \ref{fig:Fig3}b, left panel), the net magnetization is very large and parallel to the applied magnetic field, corresponding to an orbital paramagnetic response of the island.  When an extra electron is added or removed from this situation (figure \ref{fig:Fig3}b, right panel), the total magnetization is small and antiparallel to the applied field, so that the island behaves diamagnetically.  The total magnetization is roughly given by the magnetic moment of the highest singly occupied Kramers doublet. 
 
\begin{figure*}
\epsfig{file=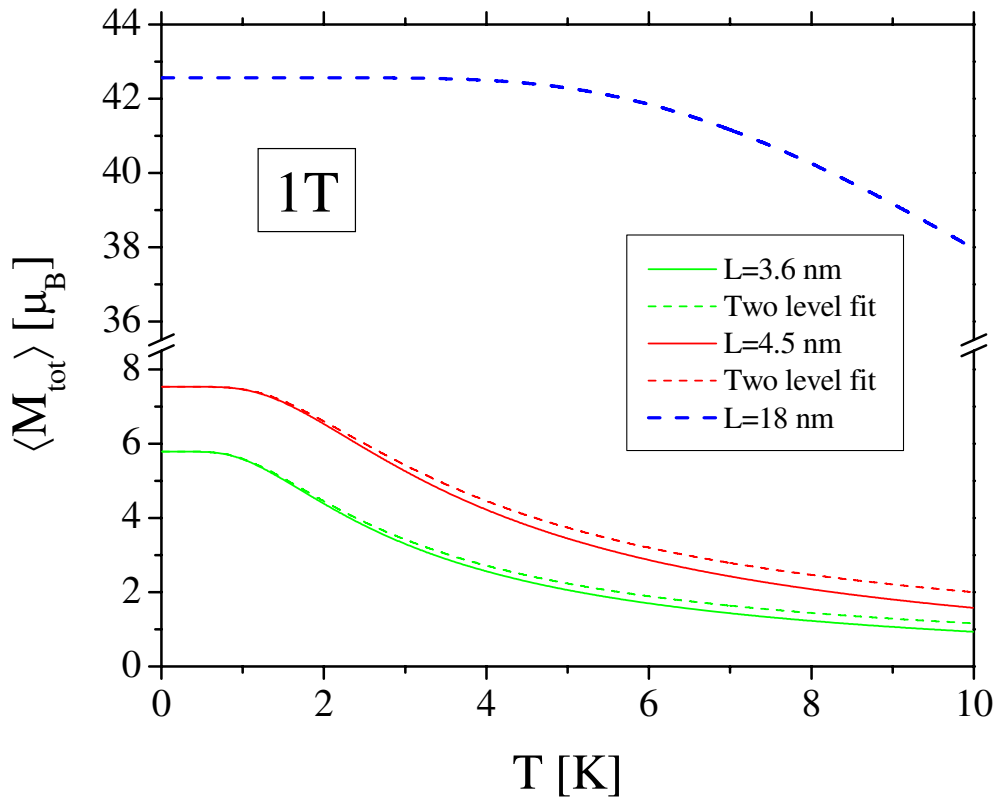,width=5.5in}
\caption{\label{fig:Fig4}{\bf Temperature dependence of total magnetization.} Temperature dependence of total magnetization $M_{tot}$ for systems with different sizes, $L\simeq 3.6$nm (green lines), $L\simeq 4.5$nm (red lines) and $L\simeq 18$nm (blue line), for a magnetic field $B=1$T and chemical potential in the middle of two states from Kramers doublet (see left panel in Fig. \ref{fig:Fig3}b). Solid lines correspond to results obtained using eq. (\ref{Magnet}) and dash lines to two-level approximate model given by eq. (\ref{MagnTwoStates}). For zero temperature the state with orbital magnetic moment $+M_n$ contributes to total magnetization $M$. With increasing temperature, the state with opposite orbital magnetic moment $-M_n$ is populated, reducing magnetization. The blue dash line predicts magnetization of the system with edge length $L\simeq 18$ nm by using eq. (\ref{MagnTwoStates}). No change of magnetization up to $T=5$K is predicted which is related to large energy level separation $\Delta_B(B=1T)\approx 5$meV shown in Fig. \ref{fig:Fig2}(b).}
\end{figure*}

We now address the robustness of the orbital magnetization with respect to thermal disorder, assuming
 that   thermal equilibrium has been reached so that $f(\epsilon_n)=\frac{1}{exp\left[\beta(\epsilon_n-\mu)\right]+1}$, where $\beta=1/k_BT$ and $\mu$ is the chemical potential,  which we fix midway between the 
 two states of the highest occupied   split Kramers doublet.  
The evaluation of $\langle M_{\rm tot}\rangle$ using eq. (\ref{Magnet}) requires the numerical calculation of the entire spectrum, possible  only
for  sufficiently small islands, such as those shown  in Fig. \ref{fig:Fig4},  with  $L=3.6$ and $L=4.5$nm and
$\langle M_{\rm tot}(T=0)\rangle=$ 5.8 and 7.5 $\mu_B$ respectively.  
Upon heating, 
the  magnetization remains stable up to   1 Kelvin and then decays. 
The temperatures at which the magnetization decays by 50 percent are 3 and 4 Kelvin respectively.  
  Since the dominant contribution to the  magnetization comes from the highest occupied Kramers doublet, with energies $E_1, E_2=E_1+\Delta_B$ and magnetic moment $M$, 
     we expect that the magnetization will be approximately given by:
\begin{eqnarray} 
\langle M_{\rm tot}\rangle\approx M\left( f(E_1)- f(E_2)\right)=M \frac{e^{\frac{\Delta_B}{2k_BT}}-e^{-\frac{\Delta_B}{2k_BT}}}{2+e^{\frac{\Delta_B}{2k_BT}}+e^{-\frac{\Delta_B}{2k_BT}}}.
\label{MagnTwoStates}
\end{eqnarray}
 The good qualitative  agreement between the exact and the approximate curves, shown in Fig. \ref{fig:Fig4}, 
 supports the use of the approximate equation to estimate  $\langle M_{\rm tot}(T)\rangle$ for 
  larger islands, for which  numerical diagonalization is out of reach. Expectedly, larger islands have larger magnetic moments, larger energy splittings $\Delta_B$ and therefore, the magnetization is more robust with respect to thermal occupation of states with opposite magnetization.  For a Bismuth(111) hexagonal island with $L=18$nm our calculations predict an orbital magnetization at $T=4$K  as large of 42$\mu_B$, that would only be depleted by 10 percent at $T=10$K.  

Two energy scales determine whether is possible to selectively occupy a single state in a Kramers doublet:   the   energy splitting between different Kramers doublets $\delta$  and, within a given Kramers doublet, the magnetic splitting $\Delta_B$.    
  As we show in Fig. \ref{fig:Fig2}b,  $\delta\propto L^{-1}$, which again reflects both  the edge character of these states as well their Dirac nature, whereas $\Delta_B\propto L$, as expected from the linear scaling of $M_n$ with $ L$, shown in Fig.  \ref{fig:Fig2}a.     So,  increasing the size of the islands makes the magnetic moment of individual edge states larger, but eventually makes it impossible to prevent  scrambling of Kramers doublets. Therefore, there is a magnetic field dependent  optimal size $\delta(L)\simeq\Delta_B(L)$ for which  orbital magnetization is maximal. We note that maximum orbital moment as a function of applied magnetic field is stable as long as there is no crossing with other states, at values of B so high than the magnetic splitting $\Delta_B$ is larger than the zero field splitting $\delta$ (see Supplementary Materials). In that case, two states of different Kramers doublets could anticross resulting in a drop of their orbital moment.

The robustness of edge spin currents in QSHI is due, ultimately, to time reversal symmetry\cite{Kane+05}. Time reversal symmetric perturbations can not produce  elastic  edge backscattering.  Application of a magnetic field breaks time reversal symmetry, which  combined with a time reversal symmetric disorder potential could, in principle, produce backscattering, resulting in the destruction of the orbital magnetization by mixing states with opposite orbital magnetization. Therefore, we test  the robustness of our predictions by studying the effect of  disorder.  We first consider Anderson disorder, introducing a uniformly distributed random potential on every orbital of the system.  This  introduces both atom to atom variations, the conventional Anderson disorder, but also random crystal field splittings at every atom.   

\begin{figure*}
\epsfig{file=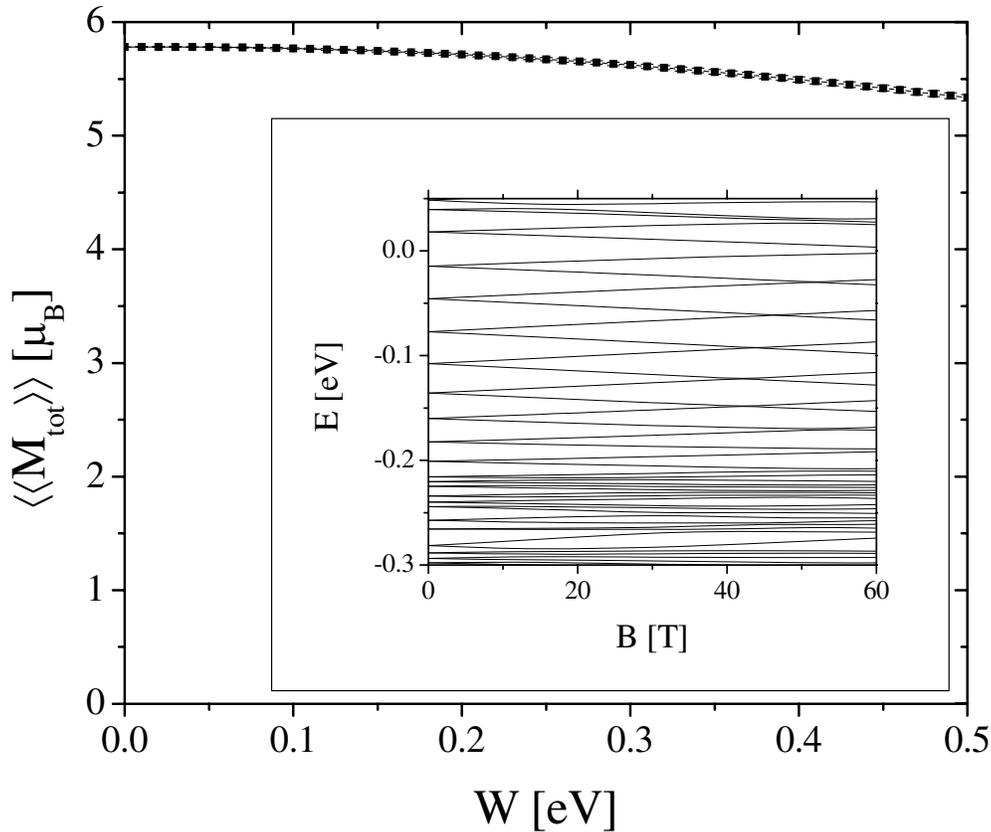,width=5.5in}
\caption{\label{fig:Fig5}{\bf Robustness of magnetization with respect  disorder.} Stability of magnetization $M_{tot}$ against disorder $W$ at temperature $T=0$ for a nanoisland with edge length $L\simeq 3.6$ nm. A random onsite energy per atomic orbital $\alpha$ is chosen from an energy interval $E_\alpha=(-W/2,W/2)$. Magnetization $M$ decreases by less than 10$\%$ for the strongest disorder, $W=0.5$. The plot includes also statistical standard deviation bars, however due to their small values they overlap with data points. The inset shows evolution of energy spectrum in a magnetic field $B$ for disordered system with $W=0.2$ eV. Dispersion of edge states in a magnetic field $B$ is not significantly different in comparison to clean system, shown in Fig. \ref{fig:Fig1}.}
\end{figure*}

The average $T=0$ total magnetization for $B=1$T for the island considered in Figs. \ref{fig:Fig1} and \ref{fig:Fig3},  obtained after 
 after averaging over 100 realizations of disorder configurations, with on-site energies randomly distributed on the interval $\pm W/2$,  denoted by $\langle\langle M_{\rm tot}\rangle\rangle(T=0)$ is shown in Fig. \ref{fig:Fig5}.  We also plot the statistical standard deviation, but  due to its small size it is smaller than the data points. 
 The stability of the orbital magnetization is remarkable even for disorder strength of $W=0.5$ eV per atomic orbital.   
 The very week effect of disorder on the edge magnetization  can also be seen (inset of Fig. \ref{fig:Fig5})  in the evolution of the in-gap edge states spectrum as a function of the applied field $B$.  
  
 In addition, we have verified that positional disorder at the edges (see supporting information Fig. S1c) does not reduce the orbital magnetization either.   We have also verified that the shape of the island   and the type of edge play no role:   similar results are obtained for  triangular zigzag islands and for hexagonal armchair islands (see supporting information).

We have also considered the influence of the substrate, relevant for the case of Bi(111) flakes on top of Bi(111) bulk\cite{Drozdov+14}. For that matter we have calculated the orbital magnetization of the edge states of zigzag hexagonal islands on top of a much larger Bi(111) flake.   Our results show that the orbital magnetization of many edge states is still preserved  in the supported islands(see supporting information).    

The phenomenon of robust orbital magnetization in nanoislands of QSHI can be also analyzed from a different perspective: these systems behave like   mesoscopic  quantum rings fabricated to observe persistent currents. 
  These  QSHI nanoislands   have three  major advantages, compared  with the conventional  quantum rings.  First, 
  there is no need to pattern any structure, since the bulk of QSHI islands is not conducting.  Second, the topological protection of the edge current flow, which our calculations show is  preserved in part in the presence of magnetic fields and disorder, results in a robust persistent current. Third, the Dirac nature of the quasiparticles permits to upscale the resulting magnetic moment linearly with size, unlike Schrodinger quasiparticles, for which the magnetic moment does not depend on size. 
    Actually,  the existence of edge states with  persistent charge current was also predicted for QSHI quantum dot made of HgTe quantum well with inverted band structures\cite{Chang+11}.
We have also verified (see Fig. S3 in Supp. Mat.) that  robust orbital nanomagnetism is present in the Kane-Mele model for a nanoisland. Our results, together with those of reference\cite{Chang+11},  
 suggest    that the physics discussed here  is universal, model independent, and thereby can be expected from any QSHI nanoisland. 
  
Experimental work will determine if the orbital magnetization will present remanence, as in the case of  nanomagnets, or at zero field the orbital moment will present random telegraph noise as in super paramagnetic particles. 
  This will be related to the fascinating question of spin relaxation between two states in a Kramers doublet, that entails a rather large change in orbital angular momentum\cite{Chudnosky+80}, and coupling to other spin degrees of freedom, such as the Bi nuclear spins,  will certainly play a role.   
  The selective occupation of a spin flavor at $B=0$, and the resulting orbital magnetization, could also be driven by Coulomb interactions\cite{Soriano+10} that we have neglected in this work.
Finally, it has not escaped our attention that  these islands could operate as well as very good interface for  single spin readout:  injection of a single electron
in an otherwise closed shell configuration will result in a orbital magnetization conditioned to the spin orientation of the added electron.  
The magnitude of the orbital magnetization would be well within the reach of state of the art local probe for magnetization, such as magnetic resonance force microscopy\cite{Rugar04} and NV center nanomagnetometry\cite{Yacoby13}.

{\it Acknowledgment}
PP thanks financial support from the Polish Ministry of Science and Higher Education, 'Mobilnosc plus' nr 1108/MOB/13/2014/0. JFR acknowledges  funding from  MEC-Spain ( FIS2013-47328-C2-2-P ) 
  and Generalitat Valenciana (ACOMP/2010/070 and  Prometeo). We acknowledge fruitful discussions with Juan Jose Palacios.  This work has been financially supported in part by FEDER funds.

\end{document}


\maketitle

\section{Supporting Information}

 \subsubsection{The edge type and nanoisland shape dependence}
\begin{figure}
\epsfig{file=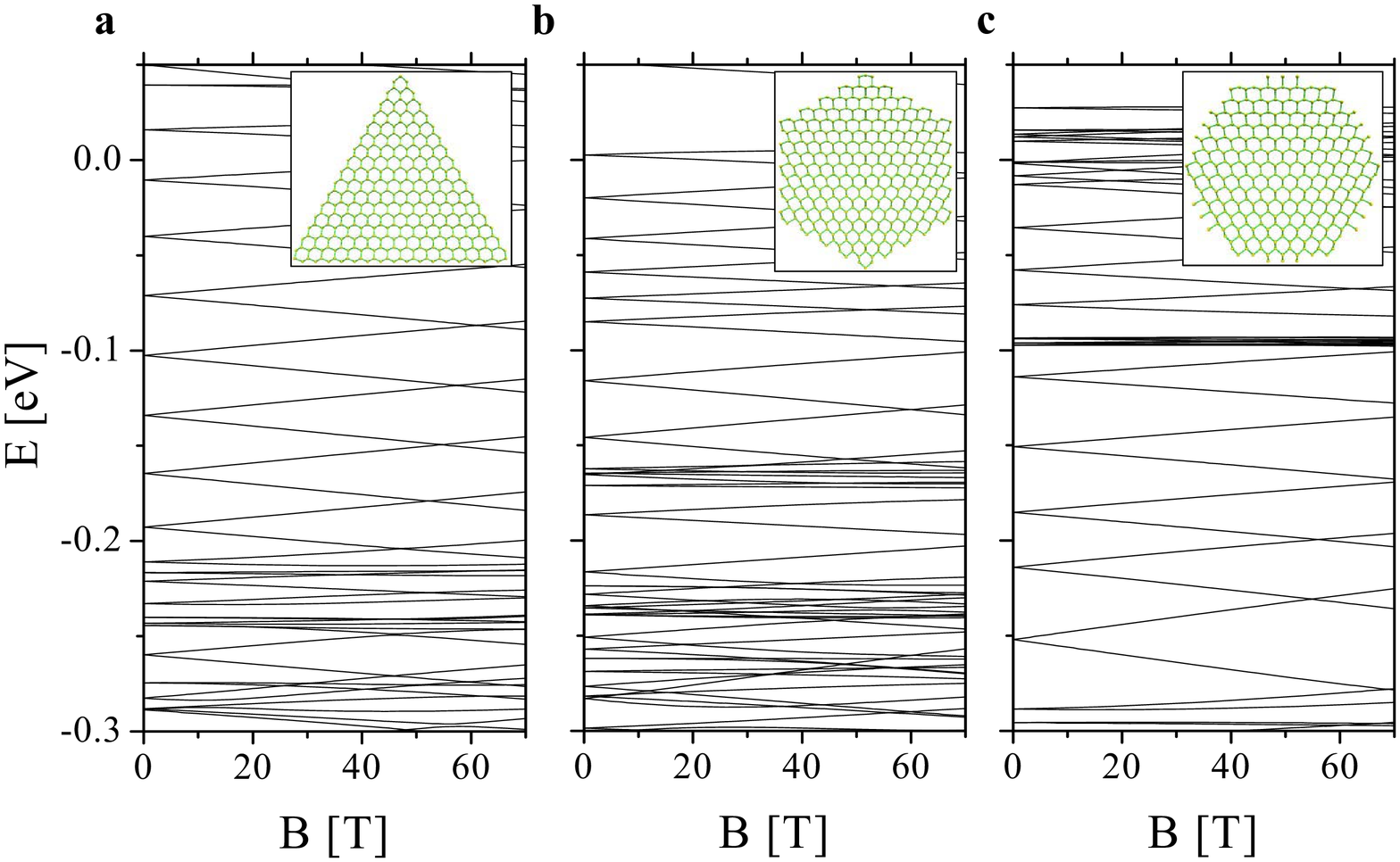,width=5.5in}
\caption{\label{fig:FigS1}{\bf Shape effect.} An evolution of energy spectra in a magnetic field of {\bf a} triangular nanoisland with zigzag edges, {\bf b} hexagonal  nanoisland with armchair edges, and {\bf c} hexagonal nanoisland with  structural edge disorder. For all structures some set of states reveal linear splitting in a magnetic field.}
\end{figure}

We  have verified that the orbital magnetization occurs regardless of the edge type and nanoisland shape.
In Fig. \ref{fig:FigS1} we show an evolution of energy spectra in a magnetic field of {\bf a} triangular nanoisland with zigzag edges, {\bf b} hexagonal nanoisland  with armchair edges, {\bf c} hexagonal nanoisland with  structural edge disorder.  All these spectra are calculated using the same four-orbital  tight-binding model discussed in the text\cite{Liu+95}.  Whereas the distribution of level spacings $\delta$  of the islands with  armchair edge and the one with a disordered edge is less regular than in the case of  islands with ideal zigzag edges,  all of them show a strong orbital magnetization, as reflected by the large splitting induced by a magnetic field. Results presented in Fig. \ref{fig:FigS1} are in agreement with an analysis from Ref. \cite{Yazyev+12} regarding an existence of topologically protected edge states for structures with arbitrary edge orientation.
 For the island with dangling atoms a linear splitting in a magnetic field for many in-gap states is still observed. These calculations  confirm the presence of orbital magnetization in arbitrary shape and type of edges QSHI nanoislands. 

\subsubsection{Influence of the substrate}

\begin{figure}
\epsfig{file=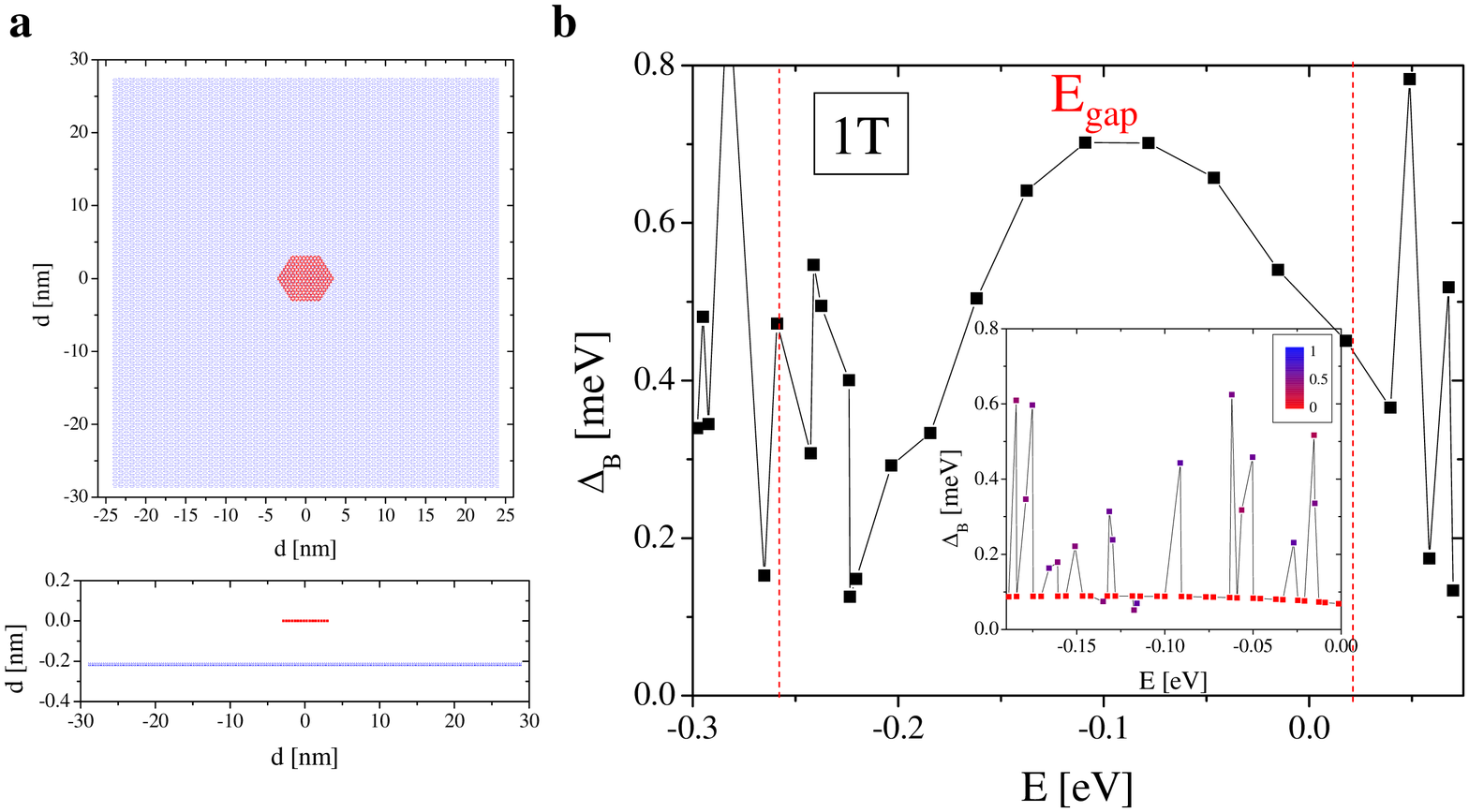,width=5.5in}
\caption{\label{fig:FigS2}{\bf Interaction with substrate.} {\bf a} Nanoisland with $L\simeq 3.6$nm ($N=384$ atoms) deposited on $50\times 55$nm ($N=31240$ atoms) substrate from top(upper panel) and side(lower panel) view. In order to reduce edge effects from the substrate, periodic boundary conditions in one direction were applied. Energy splitting of Kramer's degenerate pairs $\Delta E_B$ for a magnetic field $B=1$T without the substrate is shown in a main panel and with the substrate in the inset. The Bi (111) bilayer energy gap $E_{gap}$ is indicated by red dash lines. The states localized within nanoisland are found by calculating electronic probability densities for each energy eigenstate. States fully localized in the substrate are indicated by red color while states fully localized in nanoisland by blue color.}
\end{figure}

We now address the influence of the interaction with the substrate on the orbital magnetization. Motivated by  the experiments by Drozdov and coworkers\cite{Drozdov+14}, where  the Bi(111) flakes were located on the surface of a thin film of Bi(111),
we compute the spectrum for an  hexagonal  nanoisland such as the one in  the main text, with 
 $L\simeq 3.6$nm ($N=384$ atoms),  on top of a   much larger Bi(111) island with dimensions  $50\times 55$nm ($N=31240$ atoms).  Periodic boundary conditions are assumed along one direction (see Fig. \ref{fig:FigS2}{\bf a}) so that the substrate edges are disconnected.
   Full diagonalization of Hamiltonian matrix of such a big system (in this case size of Hilbert space $\sim 3*10^5$) is not possible, so we find only a set of eigenstates in a vicinity of bilayer energy gap $E_{gap}$ using iterative eigensolver method\cite{primme}. 
The energy splittings for  $B=1T$,  both with (inset) and without (main panel) coupling to the substrate,   are shown in Fig. \ref{fig:FigS2}{\bf b}. We focus on  the  energy range given by the in-gap region, taking 
the gap $E_{gap}$, from the  two dimensional Bi (111) bilayer,   which  is marked by vertical red dash lines.  
We find additional in-gap states of the supported island, compared to the freestanding one,   that correspond to substrate edge states, as inferred from inspection of their wave functions. In the inset we assign a color code to the states: red stands for substrate states, blue for island states. Whereas the evolution of the magnetic moments as a function of energy is no longer a smooth function, it is still apparent that many in-gap states retain their orbital magnetization.  The substrate in-gap states have a very small  splitting (red points). Thus, our calculations suggest that  orbital edge currents could still be present in Bi(111) flakes deposited on Bi(111).   We have also verified that similar results hold for larger islands, with $L\simeq 9$ nm edge length ($N=2400$ atoms) deposited on the same substrate. 

\subsubsection{A magnetic field dependence}
\begin{figure}
\epsfig{file=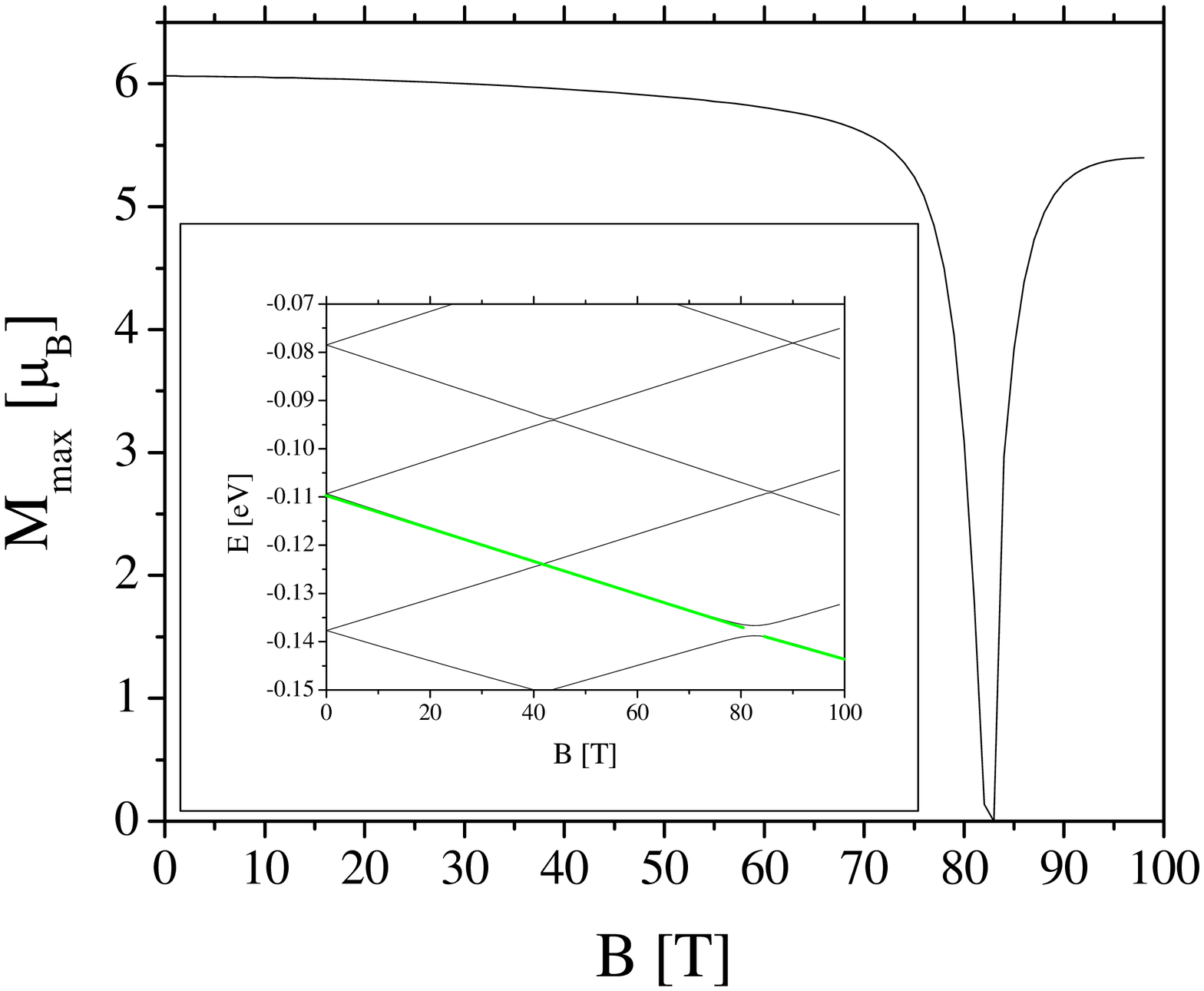,width=5.5in}
\caption{\label{fig:FigS4}{\bf Magnetic field dependence.} {\bf a} Maximum orbital moment $M_{max}$ as a function of applied magnetic field for nanoisland with $L\simeq 3.6$nm ($N=384$ atoms). The inset shows an evolution of energy levels as a function of a magnetic field. A green line indicates an energy level generating $M_{max}$.}
\end{figure}

A magnitude of maximum orbital moment as a function of applied magnetic field is investigated in Fig. \ref{fig:FigS4}. The inset shows a corresponding energy levels evolution in a magnetic field (a green line indicates a state generating $M_{max}$). The magnetic moment is stable, only slightly decreasing, up to $B=70$T. When a crossing of a given state with other state occurs, e.g. for $B=40$T, see the inset, there is no change of orbital moment due to no change of a slope of energy level in $B$. On the other hand, when two states anticross, e.g. for $B=80$T, it results in a drop of the orbital moment. When anticrossing region is left, magnetic moment return to its high value.   

 \subsubsection{Orbital nanomagnets with other Hamiltonians}
        
\begin{figure}
\epsfig{file=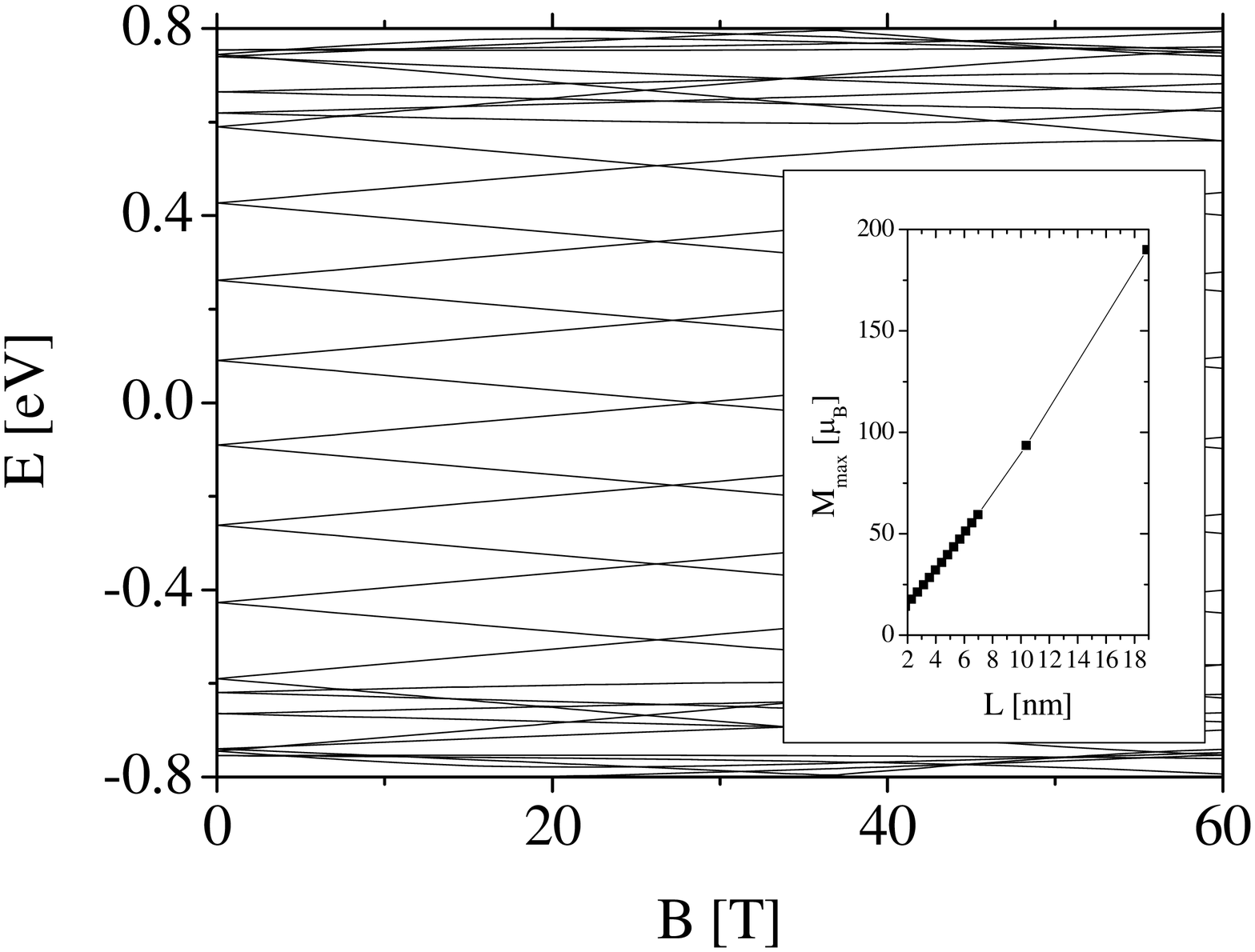,width=5.5in}
\caption{\label{fig:FigS3}{\bf Nanoisland within Kane-Mele model.} An evolution of energy spectra in a magnetic field of hexagonal nanoisland with armchair edges with $N=1986$ atoms and $L=4.4$nm calculated within Kane-Mele model for $t=-3.0$eV and $\lambda_{SO}=10$meV. In a vicinity of the middle of the energy spectrum, $E=0$ a set of linearly dispersed states in a magnetic field is clearly seen. The inset shows a linear dependence of maximal orbital magnetic moment $M_{max}$ as a function of nanoisland edge length $L$.}
\end{figure}

In order to check universality of orbital nanomagnetism in QSHI nanoislands, we model them  with  the Kane-Mele Hamiltonian \cite{Kane+05}. This model, widely used to describe the QSH phase, is a good approximation for materials such as graphene and Silicene\cite{Cahangirov+09}. It features  a single orbital per atom and the spin-orbit interaction is given by a  spin-dependent second neighbor hopping. In the presence of an external magnetic field the Hamiltonian is written as 
\begin{eqnarray}
H_{KM}= t\sum_{\left\langle i,j\right\rangle,\sigma}e^{i\phi_{ij}} a^\dagger_{i\sigma}a_{j\sigma}+i\lambda_{SO}\sum_{\left\langle\langle i,j\right\rangle\rangle\alpha\beta}v_{ij}s^{z}_{\alpha\beta} a^\dagger_{i\alpha}a_{j\beta},
\label{HKM}
\end{eqnarray}
with summation in the first term over nearest neighbors and in the second term over next nearest neighbors, and $v_{ij}=\pm 1$ for clockwise and counterclockwise direction of path connecting site $i$ and $j$, $s^{z}$ is a Pauli matrix and $\lambda_{SO}$ strength of effective spin-orbit coupling. We  have simulated a hexagonal nanoisland with armchair edges consisting of $N=1986$ atomic sites and $L=4.4$nm. In Fig. \ref{fig:FigS3} the  evolution of the energy spectrum as a function of magnetic field is shown. Again, our calculations yield a linear splitting of Kramers doublets with a magnetic field around energy $E=0$.  Using the same argumentation of the main text, this should lead to the generation of persistent edge currents whenever a single electron occupies the highest occupied Kramers doublet. In the inset, we show the size-scaling of maximal magnetic moment $M_{max}$ from these states, and we find a linear scaling, consistent with  the behavior expected for Dirac electrons in a ring. 

Finally, we note that quantum dots described with the Hamiltonian proposed by  Bernevig-Hughes-Zhang (BHZ) to predict the  QSHI phase  CdTe/HgTe quantum wells\cite{BHZ1}, as well as in  III-V type II structures\cite{BHZ2},  also lead to the existence of in-gap edge states \cite{Chang+11} in the dots endowed with orbital magnetization.   Therefore,  the concept of orbital nanomagnet built  with zero dimensional nanostructures of quantum Hall insulators is a model-independent prediction.